\documentclass[sigconf]{acmart}
\usepackage{url}
\usepackage{array}

\setcopyright{none} % No copyright notice required for submissions
\acmConference[Anonymous Submission to ACM CCS 2017]{ACM Conference on Computer and Communications Security}{Due 19 May 2017}{Dallas, Texas}
\acmYear{2017}

\begin{document}
\title{Using AI to Hack IA: A New Stealthy Spyware Against Voice Assistance Functions in Smart Phones} % TODO: replace with your title

\author{Ronjunchen Zhang}
% \authornote{Dr.~Trovato insisted his name be first.}
\affiliation{%
  \institution{Swinburne University of Technology}
}
\email{101950041@student.swin.edu.au}

\author{Xiao Chen}
% \authornote{Dr.~Trovato insisted his name be first.}
\affiliation{%
  \institution{Deakin University}
}
\email{x.chen@deakin.edu.au}

\author{Jianchao Lu}
% \authornote{Dr.~Trovato insisted his name be first.}
\affiliation{%
  \institution{Macquarie University}
}
\email{jianchao.lu@mq.edu.au}

\author{Sheng Wen}
% \authornote{Dr.~Trovato insisted his name be first.}
\affiliation{%
  \institution{Swinburne University of Technology}
}
\email{swen@swin.edu.au}

\author{Surya Nepal}
% \authornote{Dr.~Trovato insisted his name be first.}
\affiliation{%
  \institution{Data61 CSIRO, Australia}
}
\email{Surya.Nepal@data61.csiro.au}

\author{Yang Xiang}
% \authornote{Dr.~Trovato insisted his name be first.}
\affiliation{%
  \institution{Swinburne University of Technology}
}
\email{yxiang@swin.edu.au}
% \author[4]{Sheng Wen \\ Swinburne University of Technology \\ \href{mailto:swen@swin.edu.au}{swen@swin.edu.au}}
% \author[5]{Surya Nepal \\ Data61 CSIRO, Australia \\ \href{mailto:Surya.Nepal@data61.csiro.au}{Surya.Nepal@data61.csiro.au}}
% \author[6]{Yang Xiang \\ Swinburne University of Technology \\ \href{mailto:yxiang@swin.edu.au}{yxiang@swin.edu.au}}

\begin{abstract}
Intelligent Personal Assistant (IA), also known as Voice Assistant (VA), has become increasingly popular as a human-computer interaction mechanism. Most smartphones have built-in voice assistants that are granted high privilege, which is able to access system resources and private information. Thus, once the voice assistants are exploited by attackers, they become the stepping stones for the attackers to hack into the smartphones. Prior work shows that the voice assistant can be activated by inter-component communication mechanism, through an official Android API. However, this attack method is only effective on Google Assistant, which is the official voice assistant developed by Google. Voice assistants in other operating systems, even custom Android systems, cannot be activated by this mechanism. Prior work also shows that the attacking voice commands can be inaudible, but it requires additional instruments to launch the attack, making it unrealistic for real-world attack. We propose an attacking framework, which records the activation voice of the user, and launch the attack by playing the activation voice and attack commands via the built-in speaker. An intelligent stealthy module is designed to decide on the suitable occasion to launch the attack, preventing the attack being noticed by the user. We demonstrate proof-of-concept attacks on Google Assistant, showing the feasibility and stealthiness of the proposed attack scheme. We suggest to revise the activation logic of voice assistant to be resilient to the speaker based attack.
\end{abstract}

% TODO: replace this section with code generated by the tool at https://dl.acm.org/ccs.cfm

%\ccsdesc{Security and privacy}
% -- end of section to replace with generated code

 % TODO: replace with your keywords

\maketitle

\section{Introduction}
%\subsection{Background}
With the recent boost in Artificial Intelligence (AI) and Speech Recognition (SR) technologies, the Voice Assistant (VA), also known as the Intelligent personal Assistant (IA), has become increasingly popular as a human-computer interaction mechanism. Voice assistants, such as Amazon Echo\&Alexa, Samsung Bixby, Google Assistant, and Apple Siri , are widely adopted in smart devices and smartphones\cite{Monkey2017,Bixby2018,YourVA2014,Siri2011}. Instead of interacting with the smartphones by touching the screen, users can send the voice commands to activate the voice assistant on smartphones, and ask him to perform the tasks, such as sending text messages, browsing the Internet, playing music and videos, and so on \cite{VAUseage2016}. 

While voice assistants bring convenience to our daily lives, it also offers a backdoor for hackers to hack into our smartphones. Voice assistant on smartphones are usually granted high privileges to access various apps and system services, thus it can be utilised as a weapon to launch attacks to our smartphones, such as to unlock smartphones without fingerprints or PINs, forge emails, control smart home devices, and even transfer money \cite{VAFunctions}. For example, after activating the Google Assistant with the keywords ``OK Google'', you can send money to your contacts through Google Pay with the command such as ``send Bob \$50 for the dinner last night.'' 

Prior works\cite{YourVA2014,Monkey2017,HiddenVC2016,DolphinAttack2017} have demonstrated the success of attacking the smartphone via the voice assistant. The process usually includes two stage. In the first stage, the attacker stealthily records the activation voice of the smartphone owner; then in the second stage, the attacker replays the recorded activation voice to activate the voice assistant, and perform further attacks by playing pre-recorded voice commands. Though prior works showed the feasibility of such attacking model, they did not consider the awareness of the smartphone owner when launching the second stage, as the activation voice and the attacking voice commands are played via the speaker on the smartphone, there is some chance that the owner would be aware of it. In this paper, we propose a novel approach to employ Artificial Intelligent techniques to hack the voice assistants on Android smartphones. We aim to answer the following two questions: (1) how to collect the activation voice in a stealthily way; (2) how to determine a smart time to launch the attack without user's consciousness.

\begin{figure}[t!]
\center
\includegraphics[width=1\linewidth]{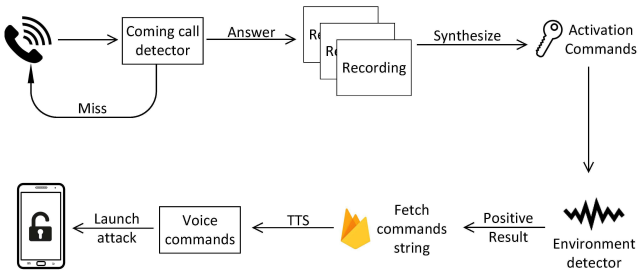}
\caption{The working flow of proposed attack model}
\label{workflow}
\end{figure}

We developed a proof-of-concept spyware app, and disguise it as a popular microphone controlled game app, which reasonably requests the permission to access the microphone. The spyware records the incoming and outgoing calls stealthily, and synthesise the activation keywords (\textit{e.g.} ``OK Google'') by adopting Nature Language Processing (NLP) techniques. After the activation keywords are synthesised, the spyware monitors and obtains the environment light and noisy data by employing the accelerometer, the ambient light sensor, and the microphone. The collected data are fed into a Machine Learning based environment recogniser to decide whether it is a optimal time to launch the attack. The working flow of the proposed attack model is demonstrated in Figure. \ref{workflow}. We tested the spyware on both industrial anti-virus products, and academic research schemes. None of them can detect the spyware.

The major contributions of this paper are summarised as follows. 
\begin{itemize}
\item We proposed an attacking model that stealthily hacks into smartphones via the built-in Voice Assistants, without the awareness of the user. 

\item We designed an intelligent environment detector to launch context-awareness attacks. We categorise the ambient environment into six real world scenarios, according to the data collected from the smartphone sensors. The environment detector determines the optimal time to launch the attack, based on analysis in current scenario.   

\item We developed a proof-of-concept spyware to practise the attack in real world. We analysed its power and computation resources consumption. The results showed that the developed spyware did not noticeably influence the performance of Android phones. We also tested it in both the industrial and academic malware detection solutions. We found that all those detection methods could not capture the new spyware.
\end{itemize}

The rest of this paper is organised as follows. Section 2 presents the background of the voice assistant. In Section 3, a novel attacking framework via built-in speaker to attack voice assistant is proposed. Proof-of-concept attacks are demonstrated. A context-awareness stealthy attacking module, together with evaluation, is presented in Section 4. The infection method and the resistance to current Anti-virus tools are presented in Section 5, along with the analysis of the power and computation resources consumption. Section 6 discusses the defence strategies, and some in-depth topics. A critical review on related works on attacks to voice assistant, and context-awareness attacks are presented in Section7. Section 8 concludes this paper.  

\section{Voice Assistant Primer}
Voice assistant is a typical application in artificial intelligence field. It is widely used for human-machine chatting, device control, and identity authentication \cite{VAEvaluation2015}. Because human beings speak around 150 words per minute but can only write 40 words per minute on average, an efficient voice recognition function will be very useful for devices like smartphones or computers to transform speeches into machine-readable texts. Currently, the voice assistant technique has been embedded into most smartphones and many apps, such as Apple Siri, Google Now, Cortana, and WeChat Voice Activation \cite{Cortana2014,AI2015,PersonalAssistant2016}. These products are very intelligent, not only recognising what human beings are saying, not also identifying who are the speakers for authentication purpose.

Existing voice assistants basically contain two functions: voice control and feedback. They can interact with human beings by receiving requests and sending feedback purely via voice just like talking with a real person. Because voice assistants usually have high authorisation in host systems, people can control the systems by inputting their sound. The first step is to deliver the activation command (\textit{e.g.}, `hello, Siri' for iPhone) to activate the system. Once the voice assistant has been activated, people can issue a request/command via voice to execute a specific function, such as making phone calls, playing music, sending messages. Our developed spyware will attack the voice assistant function in Android phones. The basic idea is to activate voice assistant through manipulated sound from the speakers on the phones themselves. The spyware can then launch a series of malicious operations through the same channel once the voice assistant has been activated. We have summarised the malicious operations that can be done once the spyware succeeds in the activation (see details in Section \ref{PoFA}).

\begin{figure}[t!]
\center
\includegraphics[width=1\linewidth]{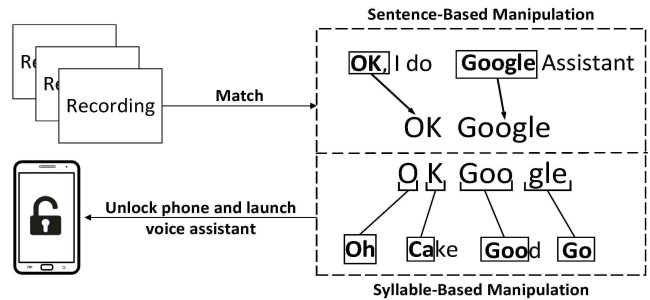}
\caption{word-based manipulation vs. syllable-based manipulation: the former requires to record the whole activation keyword,  with a high successful activation rate, while the latter only requires to record the words that contains syllables of the activation keyword, but with a lower successful activation rate}
\label{Activation_Voice_Manipulation}
\end{figure}

\section{Attack on Voice Assistants}
\subsection{Attack model}

To protect the privacy of Android users, android apps must request permissions to access sensitive user data (such as SMS and contacts), as well as system features (such as camera and Internet). Current voice assistants (such as Google Assistant and Samsung Bixby) has been granted  a great number of permissions, some of which are high risk permissions such as reading and writing private information, getting  and setting phone settings, as well as accessing system resources. Once the voice assistant is compromised, attackers can control the device via sending voice command to the voice assistant, such as making phone calls to specific numbers, sending out current geographical location via email/SMS, and switch off the Bluetooth, etc. 

The attack model include State Monitoring and Voice Recording(SMVR), Activation Voice Manipulation (AVM), Intelligent Environment Detection(IED), Voice Assistant Activation and Control(VAC). SMVR module monitors the state of the smartphone, when the microphone on the smartphone is activated (\textit{e.g.} there is an incoming/outgoing call), it starts recording the voice that the microphone receives. AVM module then process the recorded voice, and craft the activation keywords that is required to activate the voice assistant. In the IED module, we design a novel scenario recognition scheme, which collects ambient data (\textit{e.g.} light level, noisy level) through built-in sensors on the smartphone, and intelligently classify the current status into one of six finely designed real world scenarios, which determines whether it is a suitable time to launch the attack, and what is the optimal volume to play the attacking voice. If the result obtained from IED module is positive, VAC module plays the activation voice that is synthesised by AVM module, and attacking commands, in a certain volume that is determined by IED module. 

\subsection{Activation Voice Manipulation}

The built-in voice assistant can be activated either by touching and holding the home button, or by saying the activation words such as "OK Google", if the user enabled the voice activation option. The voice activation command is trained by the user's own voice to ensure that the voice assistant can only be activated by the user himself. However, besides the user saying the activation words, replaying the recorded voice of the user can also activate the voice assistant. Once the voice assistant is activated, the followed voice commands do not need to match the voice of the user. Voice commands from any sources would be executed by the voice assistant, as long as the voice command can be recognised. From the attacker's point of view,  the voice command can be pre-recorded audio files by attacker, or generated with the built-in Text To Speech (TTS) service on smartphones. Such mechanism makes the activation phase a stepping stone to control the smartphone.

The activation keywords usually consist of a greeting word (such as OK, Hi, Hey, etc.) followed by the name of the voice assistant (such as Google, Bixby, Siri, etc.) The aim of AVM module is to obtain the voice of the activation keywords that pronounced by the user. The original voice that we recorded are long sentences, so that speech recognition techniques are employed to process the voice and synthesise the activation keyword. We propose two methods for activation key synthesis, namely word-based manipulation, and syllable-based manipulation. Figure. \ref{Activation_Voice_Manipulation} illustrates the process of these two methods.

\begin{figure}[t!]
\center
\includegraphics[width=1\linewidth]{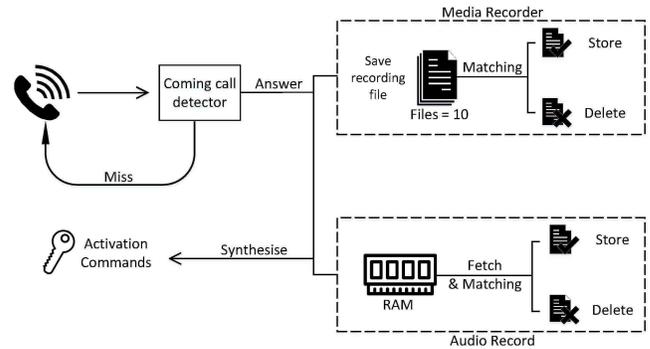}
\caption{Main different between \textsl{\textit{MediaRecorder}} and \textsc{\textit{AudioRecord}}, this two recording method will output the same result.}
\label{Examples}
\end{figure}

\begin{figure*}[t!]
\center\includegraphics[width=1\linewidth]{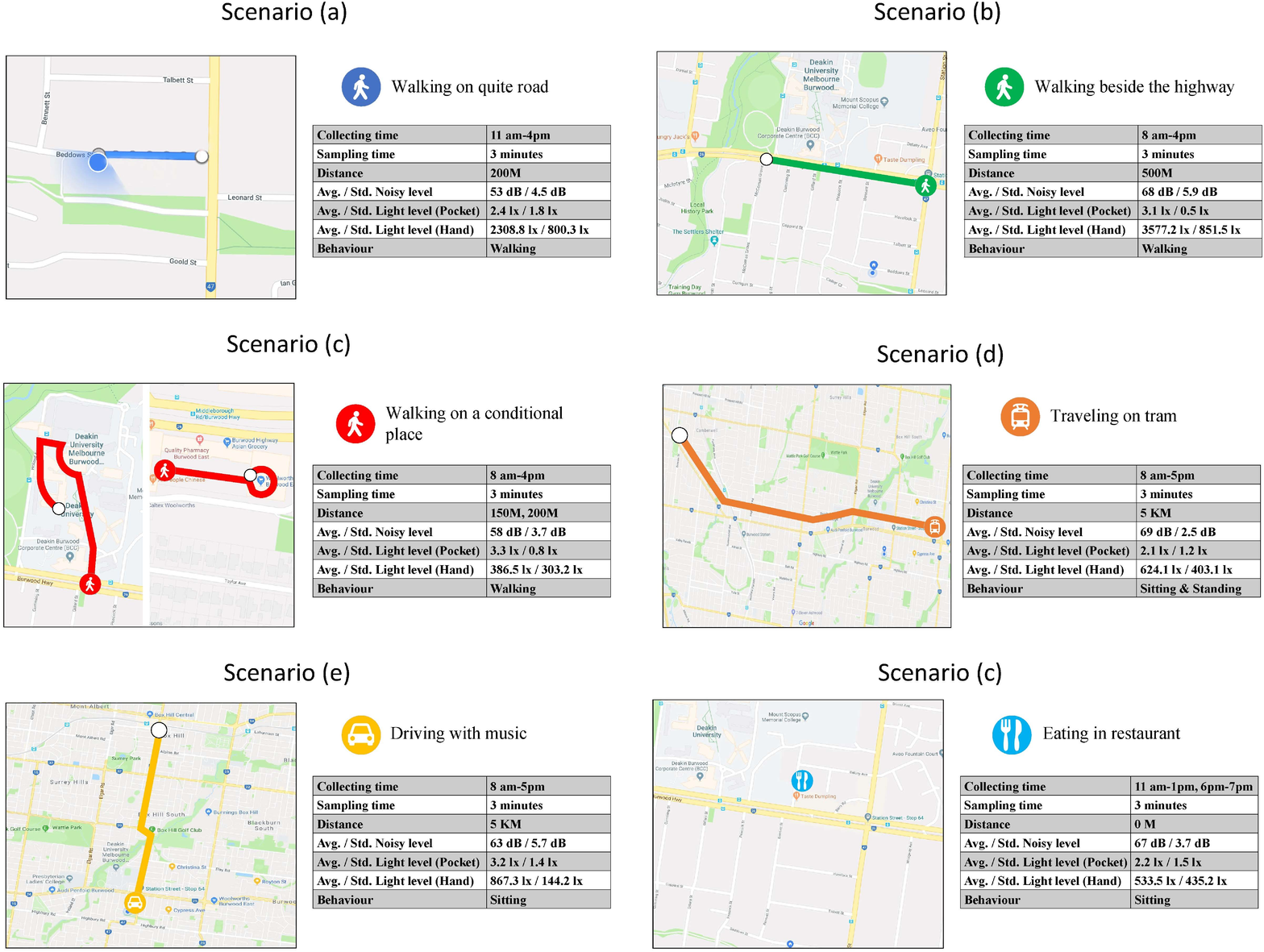}
\caption{Overview of data collected in six real-world scenarios}
\label{Scenarios_in_map}
\end{figure*}

\textbf{Word-based manipulation.}
Word-based manipulation requires to obtain the voice of the whole activation keywords. After the voices are recorded, the AVM module extracts the activation keywords and store them as audio files. Since some of the activation keywords are not usually said in the daily conversation, social engineering attacking methods can be used to improve the chance of recording the whole activation keywords. For example, in order to record the activation keyword "Google", the attacker may call the target victim and pretend to conduct a market survey, by asking "which search engine do you usually use?" The success rate of using word-based manipulation to synthesise the activation keywords achieved 100\% success rate in our experiments. 

\textbf{Syllable-based manipulation.} Collecting the whole activation word is a time consuming task, especially when the activation words contains some proper nouns, such as "Bixby". As a result we propose a syllable-based voice manipulation method, which only requires to record the syllables with the same pronunciation as the activation words. Taking the activation words "OK, Google" as an example, "OK" contain two syllables (O-K), which can be synthesised by "\underline{Oh}" and "\underline{Ca}ke". "Google" contain two syllables (Goo-gle), which can be synthesised by "\underline{Goo}d" and "\underline{Go}". When an incoming/outgoing call is detected, the AVM module records the relevant syllables and stores them in a audio file. Once each syllables are recorded, the 
AVM module syntheses the activation keywords from the collected syllables. Though Syllable-based manipulation is more flexible, it has lower successful rate compared as word-based manipulation. Approximate 40\% of the activation keywords that are synthesised by syllables can successfully activate the voice assistant. 

\subsection{Proof-of-Concept Attack}\label{PoFA}

A prototype spyware is developed and to be installed on the victim smartphone (the details about the spyware and its delivery methods are discussed in Section 5). 

\textbf{Voice recording.} The spyware registered itself as a service, which monitors the phone call status of the smartphone. Once the callback function \textsc{\textit{onCallStateChanged}}\cite{SensorEventListener} is invoked, it begins to record the audio from microphone. \textsc{\textit{MediaRecorder}}\cite{MediaRecorder} and \textsc{\textit{AudioRecord}}\cite{AudioRecord} are two classes in Android for recording audio, with \textsc{\textit{MediaRecorder}}, the audio is recorded into a file after the recording is finished, while with \textsc{\textit{AudioRecord}}, the recorded audio are cached in RAM, and APIs are provided to process the audio while recording is still in progress. We use \textsc{\textit{AudioRecord}} to achieve real-time voice synthesis. We segment the recorded audio every twenty seconds, to avoid recorded audio being overwritten. The recording process stops until the phone call is ended, or until the activation keyword "OK Google" is successfully synthesised. 

\begin{figure*}[t!]
\center\includegraphics[width=1\linewidth]{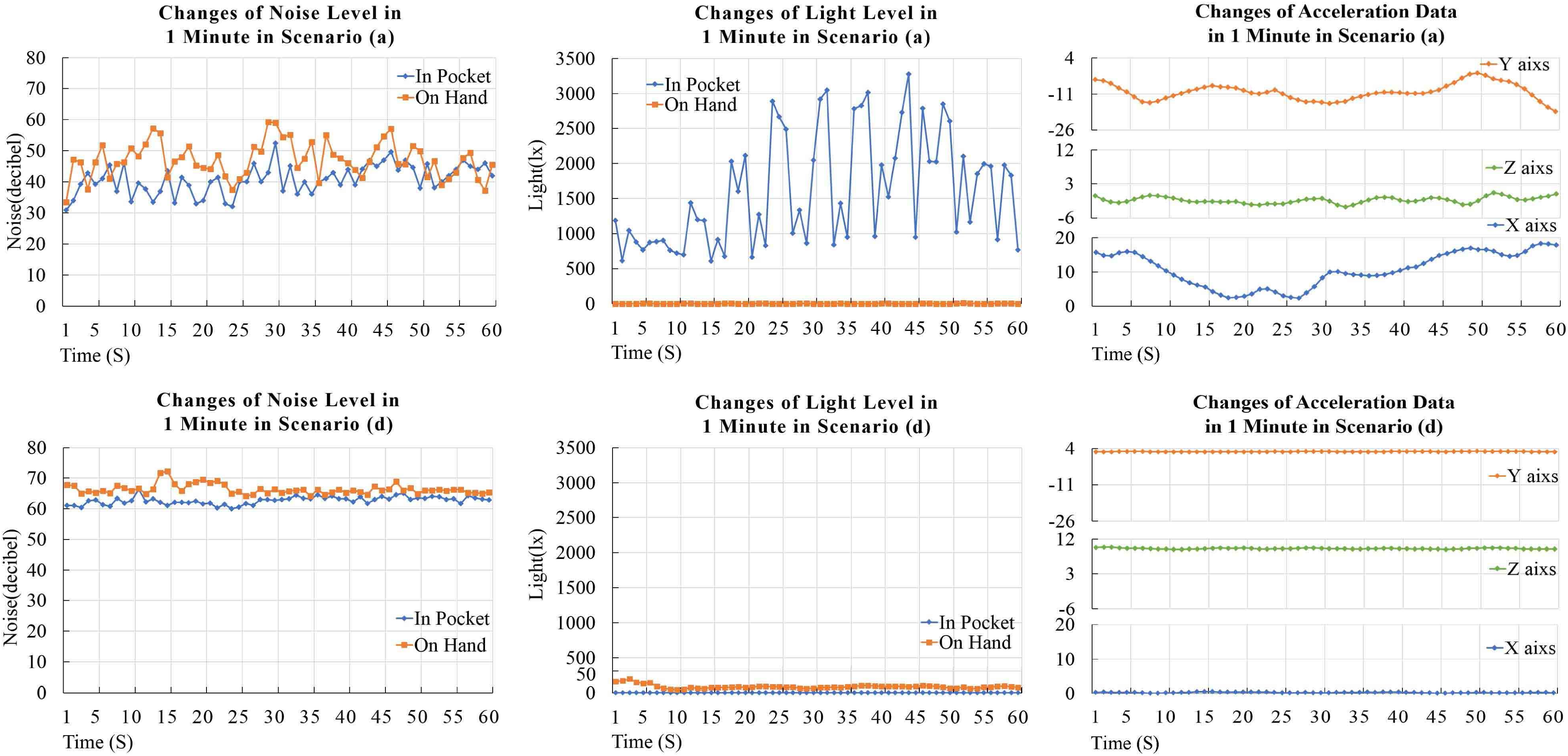}
\caption{Sample data collected from scenario (a) walking on a quiet road; and (d) taking public transportation.}
\label{Sensor_change}
\end{figure*}

\textbf{Voice recognition and synthesis.} We apply real-time speech recognition to synthesis the target activation keywords. When recording is in process, we read the recorded audio segments from the RAM, pre-processing the audio to eliminate the segments that do not include human voice, and use iFlytek speech recognition SDK to recognise and synthesise the activation keywords. Voice synthesis is working in both word-based and syllable-based modes. Once the activation keywords are synthesised, we save it as "key.wav", and validate the activation keywords by recognising it with iFlytek speech recognition SDK. If the recognised result is the same as the activation keyword, the State Monitoring and Voice Recording (SMVR) module would be disabled, and the Intelligent Environment Detection (IED) module is activated to collect environment data. 

\textbf{Environment detection.} Once the activation keywords are synthesised, the IED module starts to collect data from the microphone, the ambient light sensor, and the accelerometer. Based on the collected environment data, the IED module determines whether to launch an attack. Details of Environment detection module is discussed in Section 4.

\textbf{Launching the attack.} Once the environment detection algorithm determines to launch the attack, the synthesised activation voice would be played via the speaker on the victim smartphone. Meanwhile, the attacking commands, which are in text format, are fetched from Firebase \cite{Monkey2017}. Firebase provides a realtime database and backend as a service. The service provides an API that allows application data to be synchronized across clients and stored on Firebase's cloud. With Firebase, we are able to dynamically change the attacking commands depends on the target of the attack. Attacking commands are converted to speech using the built-in Text-To-Speech service, before played via the speaker.

\textbf{Post-attacking scenarios and capabilities.} Voice Assistants are granted with high privilege, which is able to access system resources and private information. For instance, we can use Google Assistant to send SMS and emails, take photos, open other apps, take screen shots, and even transfer money via Google Pay \cite{VAFunctions}. With voice wake function, the smartphone would be unlocked without PIN or fingerprints, once the voice assistant is activated. As a result, once the attackers obtain the activation voice of the voice assistant, they can hack into victims' smartphones, no matter whether they have the phones or not.

\section{Stealthy Attacking Module}

In this section, we present the technical details of Intelligent Environment Detection module and its triggering algorithm. 

\subsection{Challenges}

After the attacker acquires the activation voice from the user, a critical question arises and yet to be answered, that is, when is the good time to launch the attack?
Obviously, there are many factors to be considered, such as whether the user is holding or interacting with the phone, whether the sound would be noticed by the user when launching the attack, whether the sound would be "heard" by the phone when launching the attack, and so on. The question is critical because once the attack is noticed by the user, he may look into the issue, and there would not be a second chance to launch the attack. Existing works \cite{YourVA2014,Monkey2017} gather the state of the phone from sensors on smartphones (\textit{e.g.} light sensor and accelerometer) and system attributes (\textit{e.g.} current system time and screen on/off status), but the attack would only be triggered under certain conditions. For example, in \cite{YourVA2014},  attack would only be launched at the night, when the phone is screen-off and put on a horizontal table with room lamp is off.  As different users may have different habits, this kind of pre-set conditions may never be met.  
The volume of the sound that played to launch the attack is also an important factor that affects the success of attacking. The volume should be low enough to avoid being heard by the user, and high enough to make sure the attack voice command can be received by the smartphone. The optimal volume varies depends on the noise level of the surroundings as well as the real-world scenario that the phone is in. 

\subsection{Triggering Algorithm}
To tackle the above mentioned challenges, we propose an Intelligent Environment Detection module, which makes decision on when to launch the attack, and what is the optimal volume to play attack voice commands, by analysing the data collected from the ambient, together with the state of the phone. 

\textbf{Sensor set-up.} We collect data from microphone, ambient light sensor, and accelerometer. Smartphones do not have a built-in noise sensor, thus we record the ambient sound via microphone, and calculate the noise level in decibel with the following formula.
\begin{displaymath}
\mathbf{L_{dB}} = 10\,log_{10}\left(\frac{A_{1}^2}{A_{0}^2}\right)
\end{displaymath}
where $A_{1}$ is the amplitude of the recorded sound, and $A_{0}$ is a standard amplitude that we set to 1. 
To model the movement patterns, we collect the data from the accelerometer in a frequency of 50 Hz. To save to power consumption, we collect the noise and light data every 200ms, as they are more stable than the movement data in a short period of time. All of the collected data are resampled to 50Hz with Nearest Neighbour Interpolation. We collected the lock screen on/off status with the system API \textsc{\textit{PowerManager.isInteractive()}}, and bluetooth and headphone connection status with \textsc{\textit{AudioManager.getDevices()}}.

\begin{table}[h!]
    \caption{Success rate in each scenario}
    \begin{tabular}{|c|m{1.2cm}<{\centering}|m{1.7cm}<{\centering}|m{1cm}<{\centering}|}
    \hline
    &Being Noticed&Successfully Attacked&Success Rate\\
    \hline
    Scenario(a) & 20 / 20 & 0 / 20 & 0 \%  \\
    \hline
    Scenario(b) & 0 / 20 & 18 / 20 & 90 \%  \\
    \hline
    Scenario(c) & 24 / 40 & 13 / 40 & 32.5 \%  \\
    \hline
    Scenario(d) & 0 / 20 & 17 / 20 & 85 \%  \\
    \hline
    Scenario(e) & 20 / 20 & 0 / 20 & 0 \%  \\
    \hline
    Scenario(f) & 0 / 20 & 19 / 20 & 95 \%  \\
    \hline
    \end{tabular}
    \label{Success_rate}
\end{table}

\textbf{Data Collection.} We collected the training data in six typical real-world scenarios, which includes motion/stationary states, noisy/quiet states, bright/dark states: (a) walking/jogging on the quite road; (b) walking/running along the highway; (c) walking in specific places; (d) taking public transportation; (e) driving/sitting in a car; and (f) eating in restaurant. In all of the scenarios, we collected the data of putting the phone in the pocket, as well as holding the phone on hands.

We find ten volunteers to collect the data in each of the above mentioned six scenarios, with Google Pixel 2 and Samsung Galaxy S9. All of them were told the purpose of the experiment, which is to make sure their attentions are not disturbed by other environmental factors.

\begin{table}[h!]
    \centering
    \caption{"Movement intensity" Features}
    \begin{tabular}{|m{1cm}<{\centering}|m{1.2cm}<{\centering}|c|}
    \hline
    Motion State&Stationary State&Movement Intensity \\
    \hline
    0.70 & 0.30 & Definite motion state  \\
    \hline
    0.56 & 0.44 & Relative motion-stationary state  \\
    \hline
    0.85 & 0.15 & Definite stationary state  \\
    \hline
    0.45 & 0.55 & Relative motion-stationary state  \\
    \hline
    \end{tabular}
    \label{Movement intensity}
\end{table}

We prepared an activation voice that can successfully activate Google Assistant in the devices, which would be played randomly between one to five minutes after the experiments started. Once the Google Assistant was activated, attacking commands, which are in text format, would be fetched from Firebase server and played via the built-in Text-To-Speech (TTS) service. We record whether the volunteer noticed the sounds, and whether the attack commands were successfully executed by Google Assistant. Note that if the attacking voice are heard by the volunteers, we label the attack as failed even if it was successfully executed by Google Assistant. Table.\ref{Success_rate} show that the success rate in each scenario.

The data that we collected from the sensors are noise level, light level, and acceleration data in x,y,z axes.
20 groups of data were collected, each with length of 3 minute. Figure.\ref{Scenarios_in_map} shows the overview of the collected data, and Figure. \ref{Sensor_change} shows the data collected from three sensors in 1 minute.

\begin{figure}[t!]
\center\includegraphics[width=1\linewidth]{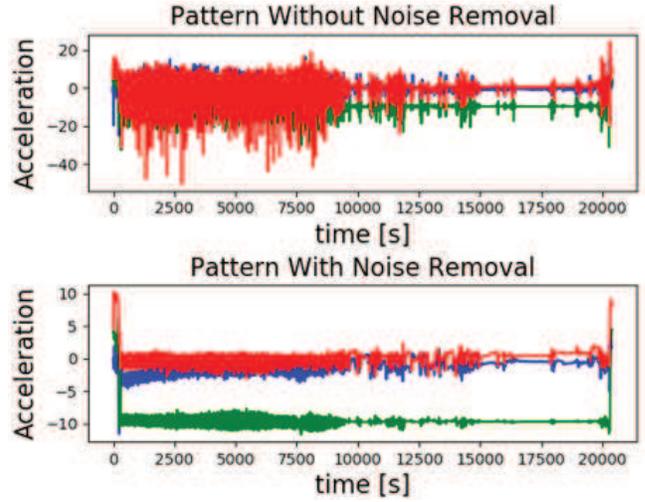}
\caption{Effect of butter wroth filter}
\label{Butter_wroth}
\end{figure}

\textbf{Noise Removal.} Raw signal usually contains noise that arises from different sources, such as sensor miscalibration, sensor errors, errors in sensor placement, or noisy environments and so on. These noisy signal adversely affect the signal segmentation, feature extraction and then significantly hamper activity prediction. In our study, we used fourth order Butterworth low-pass filter for the noise removal. Figure.\ref{Butter_wroth} compares the sensor signal pattern with and without noise removal for the same activity.

\begin{table}[h!]
    \centering
    \caption{One Hot Encode "Movement intensity" Features}
    \begin{tabular}{|c|c|}
    \hline
    Movement Intensity&One Hot Encode \\
    \hline
    Definite Motion state &  [0, 1] \\ 
    \hline
    Definite Stationary state &  [1, 0] \\
    \hline
    Relative Motion-stationary state & [1, 1]  \\
    \hline
    \end{tabular}
    \label{One Hot}
\end{table}

\begin{figure}[t!]
\center\includegraphics[width=1\linewidth]{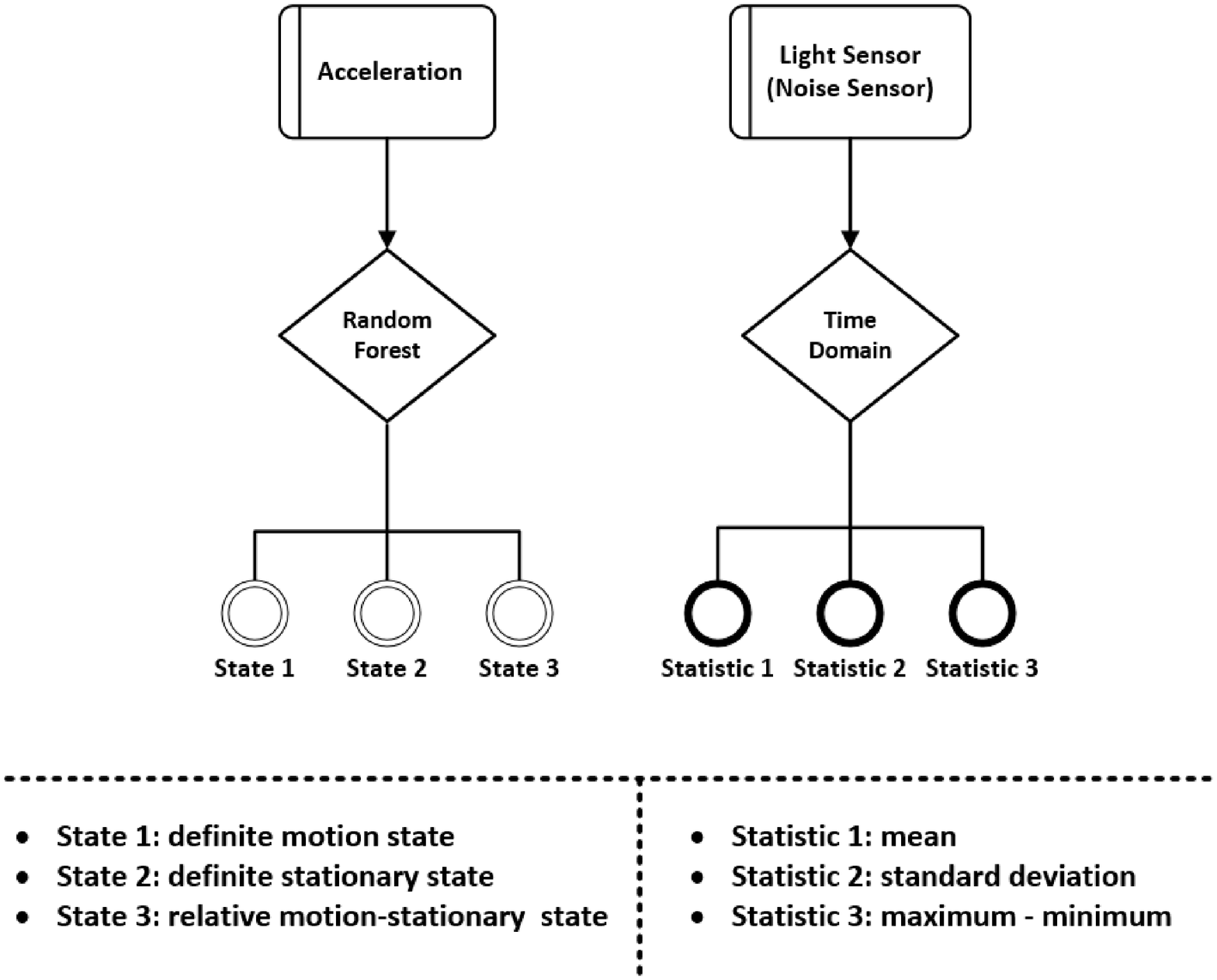}
\caption{Feature Extraction Framework.}
\label{Feature Extraction}
\end{figure}

\textbf{Feature Extraction.} Fig.\ref{Feature Extraction} illustrates the framework of proposed feature extraction approach, where the features that calculated from accelerometer as "movement intensity" and the features that extracted from noise sensor and light sensor as "environment variables".
We used "movement intensity" features to describe an overall perspective of human behaviour state. We divided human behaviour state into definite motion state, definite stationary state and relative motion-stationary state based on the class probabilities of Random Forest (RF). The RF does not directly output class labels but probabilities, and then assign labels to the instances depends on whether the probabilities exceed certain threshold. We labelled the motion state with a threshold over 60\% probability as definite motion state, the stationary state with a threshold over 60\% probability as definite stationary state, and the motion (or stationary) state with the probability between 40\% to 60\% as relative motion-stationary state (as shown in Table.\ref{Movement intensity}). As "movement intensity" features are categorical values and Machine learning algorithms cannot work with them directly, we converted all the "movement intensity" features to numerical values using One Hot Encode as shown in Table.\ref{One Hot}. Environment variables were applied for the purpose of providing more specific details on the uncertain environmental factors, such as noise level and light intensity, which can also affect the decision of whether to launch attack. 

\begin{table}[h!]
    \centering
    \caption{Average Accuracy Performance}
    \begin{tabular}{|c|c|c|c|}
    \hline
    &Precision&Recall&f1-score\\
    \hline
    Unsuccessful Invasion & 0.96 & 0.95 & 0.95  \\
    \hline
    Successful Invasion & 0.97 & 0.98 & 0.98  \\
    \hline
    Avg & 0.97 & 0.97 & 0.97  \\
    \hline
    \end{tabular}
    \label{Average Accuracy Performance}
\end{table}

\textbf{Evaluation Methods and Results.} We implement our algorithm in python 3.6, using Scikit-learn 0.17.1 package which is an open source Python library to implement machine learning algorithms \cite{Scikit2011}. We evaluate classification algorithm based on four metrics, which are True Positive rate, F1 measure, Precision, and Recall. Table. \ref{Average Accuracy Performance} shows the overall performance of Random Forest classifiers on detecting the effective attack opportunities based on the 20-fold cross validation. Table.\ref{Parameters for Random Forest} shows the parameter settings of Random Forest classifier.

\begin{table}[h!]
    \newcommand{\tabincell}[2]{\begin{tabular}{@{}#1@{}}#2\end{tabular}}
    \centering
    \caption{Parameters for Random Forest}
    \begin{tabular}{|c|c|}
    \hline
    &\tabincell{c}{Random Forest}\\
    \hline
    Parameters & \tabincell{c}{'bootstrap': True, \\ 'criterion': gini,\\'max depth': 10, \\'number of estimators': 200} \\
    \hline
    \end{tabular}
    \label{Parameters for Random Forest}
\end{table}

\textbf{Volume control.} The output activation command requires to reach a certain volume to be captured by the device. This triggering volume is affected by the environment noise level. To investigate how environment noise affects the triggering volume, we test the sufficient volume to activate the voice assistant, as well as to trigger the voice attack. As in the proposed attack scenario, the attacking voice commands are played via the speaker of the smartphone, so that the distance between the source of the sound and the device is not considered. Table. \ref{Min_voice} shows the minimal volume of activating voice assistant and triggering voice attack.

\begin{table}[h!]
    \centering
    \caption{Minimal volume for successfully activating voice assistant and triggering voice attack}
    \begin{tabular}{|m{2cm}<{\centering}|m{2.3cm}<{\centering}|m{2.3cm}<{\centering}|}
    \hline
    Ambient Noises&Activation Commands&Attacking Commands \\
    \hline
    30 dB & 44 dB & 40 dB  \\
    \hline
    41 dB & 44 dB & 40 dB  \\
    \hline
    52 dB & 54 dB & 54 dB  \\
    \hline
    68 dB & 67 dB & 67 dB  \\
    \hline
    73 dB & 75 dB & 76 dB  \\
    \hline
    \end{tabular}
    \label{Min_voice}
\end{table}

\subsection{Evaluation}

We conducted a set of real-world experiments to test the feasibility of the Intelligent Environment Detection algorithm. Ten volunteers are involved in the experiments. The experiments are conducted in different time period of a day, and in various real-world scenarios. Google Pixel 2 and Samsung Galaxy 9 are used in the experiments. In every three minutes, the IED module collects data from the sensors, and determines whether to launch the attack. If an attack is launched in the three-minute time period, the IED stops monitoring the environment, and wait for the next three-minute time period. When the IED decides to launch the attack, it first plays the activation voice, and then randomly fetch one of the twenty voice commands from the Firebase server and convert it to audio through TTS service. 

Table. \ref{IED_evaluation} shows the results of the experiments. We count the number of attack launched as if the IED module decides to launch the attack. Either the voice commands cannot be recognised by the voice assistant, or the volunteer heard the voice commands, we consider the attack is failed. In the bus scenario, 20 attacks were launched, with one attack failed, which is due to a sudden quiet when the attack command is being played, and the volunteer noticed the attacking voice. Only two and zero attacks are launched in the university and classroom scenario, respectively, because it is quiet in these two scenarios at most of the time. In restaurant scenario, 19 attacks were launched, all of which were succeeded. 12 attacks were launched in supermarket scenario, and 10 of them were succeeded. One of the two failures was caused by attack being noticed by the volunteer, and the other is because the voice assistant cannot recognise the attack command.

\begin{table*}[t!]
    \centering
    \caption{Evaluation of Intelligent Environment Detector module}
    \begin{tabular}{|c|c|c|c|c|}
    \hline
    Time Period & Activity & Number of Attack Launched & Number of Attack Succeeded & Success Rate \\
    \hline
    8:00-9:00&Sit on bus&20&19&95\%  \\
    \hline
    11:00-12:00&Walk in university&2&2&100\%  \\
    \hline
    13:00-14:00&Have lunch in restaurant&18&18&100\%  \\
    \hline
    15:00-16:00&Sit in classroom&0&0&N/A  \\
    \hline
    19:00-20:00&Shop in supermarket&12&10&83.3\%  \\
    \hline
    \end{tabular}
    \label{IED_evaluation}
\end{table*}

\section{Delivery \& Detection Avoidance}
\subsection{Spyware delivery and obfuscation.} 
We developed a proof-of-concept spyware on Android platform to launch the attack, which can be is called as infection process\cite{IEMITF2015,XN2009,HidenCode2006}. Three permissions are required for the spyware to perform the malicious activities, which are \textsc{\textit{RECORD\_AUDIO}} (to record the activation voice of the user), \textsc{\textit{INTERNET}} (to dynamically fetch attacking commands from Firebase server), and \textsc{\textit{READ\_PHONE\_STATE}} (to monitor incoming/outgoing call status). 

\begin{figure}[t!]
\center\includegraphics[width=1\linewidth]{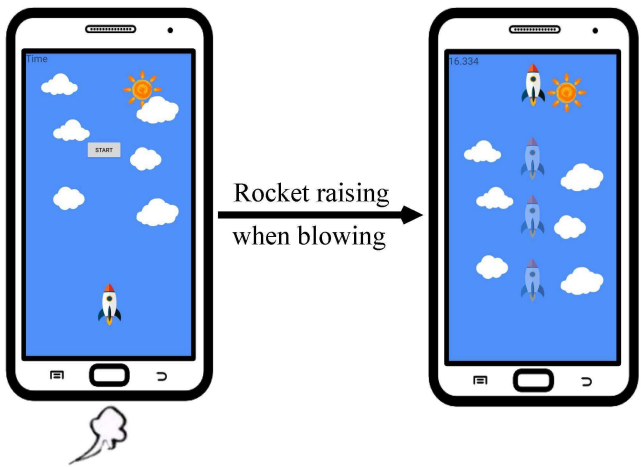}
\caption{The UI of RocketGo game. After player clicking start button, the rocket will raise when player blows or scream to the microphone. The rising speed depends on the volume of sound that the microphone receives}
\label{Infection_Game}
\end{figure}

The spyware is disguised as a popular microphone controlled game named RocketGo. The UI of RocketGo is displayed in Figure.\ref{Infection_Game}. When playing the game, the player requires to blow or scream to the microphone. The higher volume the microphone receives, the faster the rocket flies. The game is controlled by microphone, so that we can request for the \textsc{\textit{RECORD\_AUDIO}} permission without being suspected by the user. We claim in the game that game scores can be shared with friends in the contact list, so that we request the permission of \textsc{\textit{READ\_PHONE\_STATE}} and \textsc{\textit{INTERNET}}. Actually, these two permissions are very common in Android games. Five of the top10 games on Google Play requested the \textsc{\textit{READ\_PHONE\_STATE}} permission, while all of the top10 games requested the \textsc{\textit{INTERNET}} permission.

When RocketGo is launched, it registered itself as a service, which monitors the phone call state. Once the activation voice is successfully synthesis, it stops monitoring the phone call state, and start collecting environment data from the sensors to look for suitable time to launch the attack. The service would run at background even when user close the RocketGo app. User would not get any prompts in the whole process.

\subsection{Resistance to Anti-Virus Tools}
\begin{figure}[t!]
\center\includegraphics[width=1\linewidth]{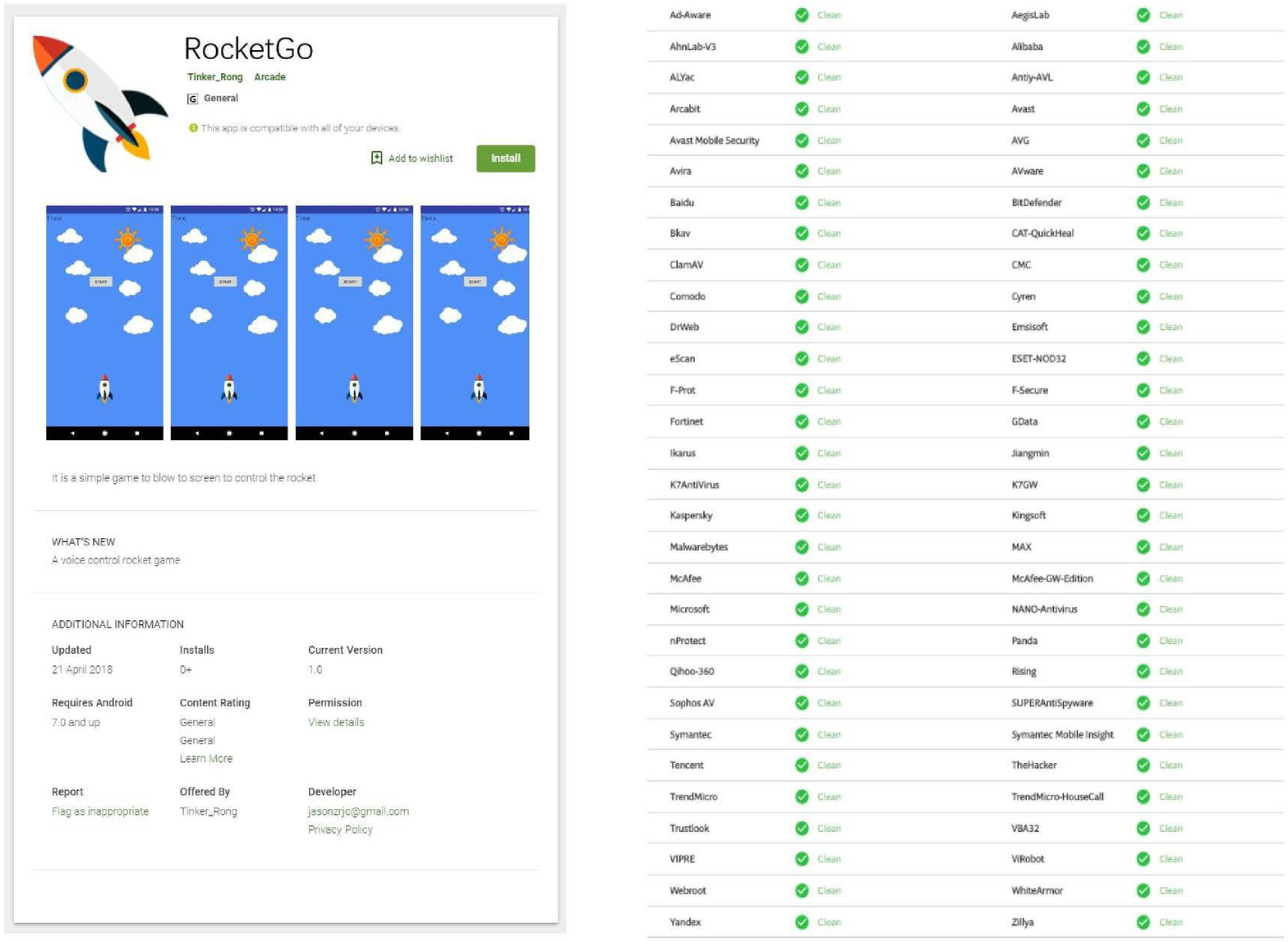}
\caption{Detection result of Google Bouncer and Virus Total}
\label{DetectionResult}
\end{figure}

We test the spyware against both industrial anti-malware tools, as well as academic malware detection solutions. 
Android malware detection approaches can be categorised into static tools and dynamic platforms, according to whether the candidate app needs to be executed or not. Static approaches are based on analysing static features of the app, such as the component of the app, the permissions requested by the app, and the code itself. Dynamic approaches execute the app in a protected environment, provide all the emulated resources it needs, and observe its malicious activities. 
For industrial anti-virus products, we test RocketGo on VirusTotal, as well as the top 10 most popular anti-virus tools on Google Play, such as Norton Security and Antivirus, Kaspersky Mobile Antivirus, McAfee Mobile Security, and so on. None of them reported our RocketGo as malicious. We also submit the RocketGo to Google Play store, where submitted apps are tested against their dynamic test platform Google Bouncer. RocketGo successfully passes the detection of Google Bouncer. The detection result of Google Bouncer and VirusTotal are present in Fig. \ref{DetectionResult}. Note that we took down the RocketGo app from the Google Play immediately after it passed the test.

For academic malware detection solutions, we test RocketGo with Drebin, which is a state-of-art machine learning based malware detection scheme\cite{Drebin2014}. We trained Drebin with 5,000 malware samples shared by the authors of Drebin, and 5,000 benign apps that we downloaded from APKPure. RocketGo is labelled as a benign app. The results show the resistance of the proposed attacking method to both industrial and academic malware detection tools.

\subsection{Consumption Analysis}
The proposed attacking process can be divided into four phases: P1(Phone call state monitoring), P2(Recording and synthesising activation command), P3(Environment monitoring), and P4(Attacking via speaker). We evaluate and analyse the power and memory consumption of each phase.

\begin{figure}[t!]
\center\includegraphics[width=1\linewidth]{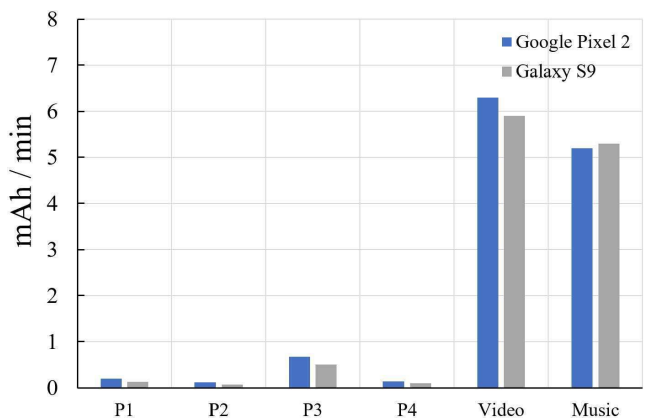}
\caption{Power consumption of four phases: P1(Phone call state monitoring), P2(Recording and synthesising activation command), P3(Environment monitoring), and P4(Attacking via speaker)}
\label{Power_consumption}
\end{figure}

\textbf{Power consumption analysis.} We install RocketGo on Google Pixel 2 (with 3520 mAh Battery) and Samsung Galaxy S9 (with 3000 mAh Battery), and test the power consumption on both of them. Fig.\ref{Power_consumption} reports the per minute power consumption of four attacking phases, compared with playing 1080P video and music. The results show that in P1, P2, and P4, the per minute power consumption on both devices are very low (approximately 0.2 mAh, 0.1 mAh, and 0.1 mAh, respectively). P3 has the highest power consumption during the whole attacking process, which is approximately 0.8 mAh per minute. It is still negligible when compared with playing video and listening to music, which consumes 6.1 mAh and 5.1 mAh per minute, respectively. We further reduce the frequency of collecting data from sensors in P3 from 50Hz to 10Hz (\textit{i.e.} 10 groups of data collected per second), the power consumption reduced to 0.5 mAh, without affecting the successful rate of attacking. The results show that the prototype spyware consumes very little power that can hardly be noticed by the user.

\begin{figure}[t!]
\center\includegraphics[width=1\linewidth]{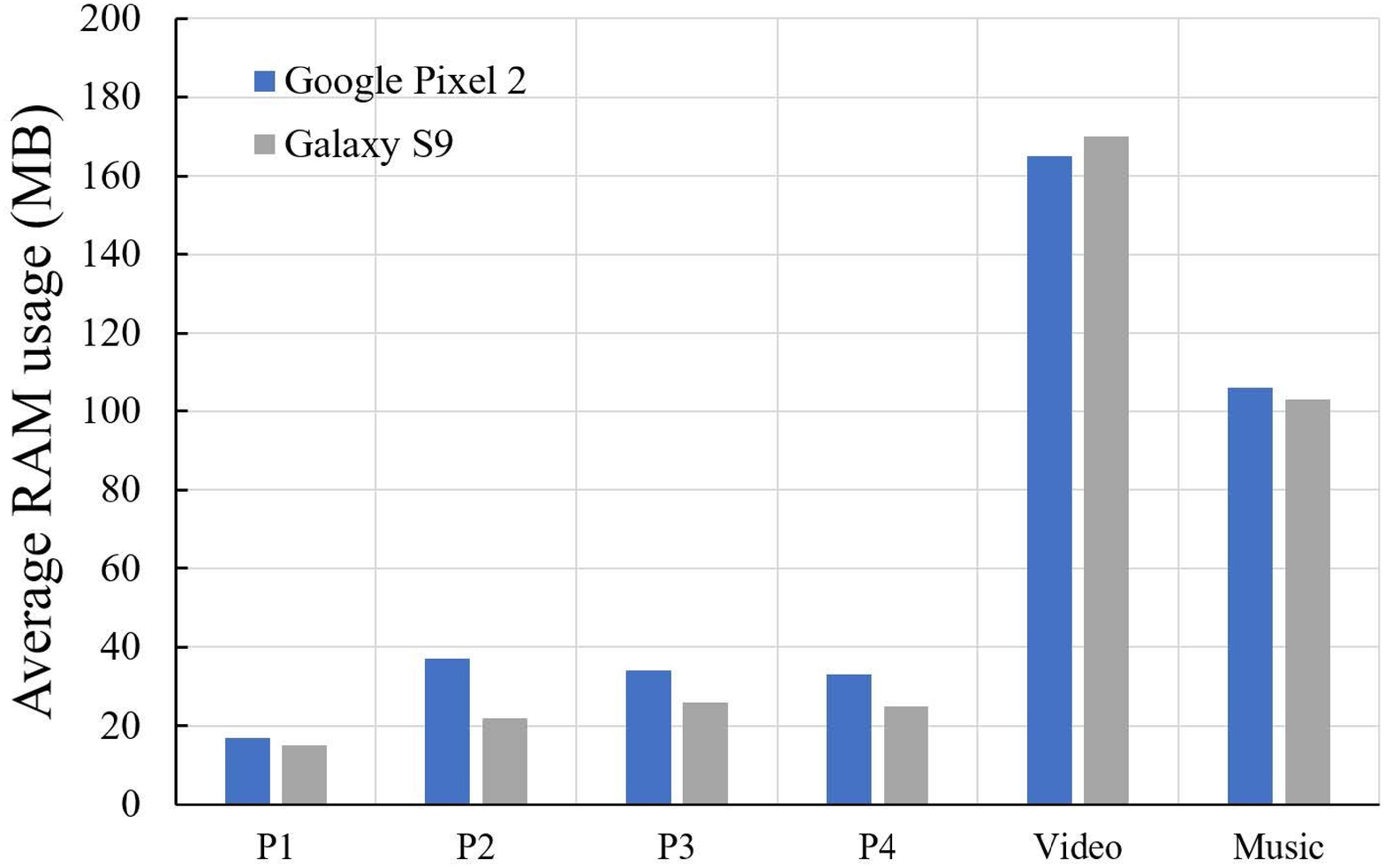}
\caption{Memory consumption of four phases: P1(Phone call state monitoring), P2(Recording and synthesising activation command), P3(Environment monitoring), and P4(Attacking via speaker)}
\label{RAM_consumption}
\end{figure}

\textbf{Memory \& CPU analysis.} We test the memory consumption of RocketGo on Google Pixel 2 and Galaxy S9, both of which has 4GB RAM. Fig.\ref{RAM_consumption} shows the average RAM consumption in the four processes. On Samsung Galaxy S9, the RAM consumption of four phases are 15MB, 22MB, 26MB, and 25MB, respectively, while on Google Pixel 2, it consumes 17MB in P1, and 35MB in P2, P3, and P4. Compared with video playing and listening to music, which consumes 170 MB and 106 MB, respectively, the memory consumption of RocketGo can hardly affects the performance of Android systems, thus it is hard to be noticed by the user.

We also evaluate the CPU consumption on Google Pixel 2 ( with 1.9GHz octa-core processor) and Galaxy S9 (with 2.8 GHz octa-core processor). It is found that only  P3 is using the CPU to compute, which consume around 7\% of total bandwidth.

\textbf{File size.} We also evaluate the size of the app in different attacking phases. During the whole attack process, two files are stored on the disk: an audio file (*.pcm) to store the synthesised activation voice, and three text files (*.txt) to record the data collected from the sensors. Table.\ref{File_size} shows the average size of each files. As we process the voice that we recorded in P2 in real-time, any recorded voices that do not contain the activation words or syllables would be abandoned, so that only the activation voice would be saved in the audio file. In P3, the data collected from the sensors is stored for X minute, and would then be overwritten by new data, to limit the size of the app. In P4, attacking commands are fetched from Firebase server in the format of texts, and convert to audio by using the built-in Text-To-Speech (TTS) services. The commands are cached in RAM, so that no file is created in this phase.  

\begin{table}[h!]
    \centering
    \caption{Average File Size (One File)}
    \begin{tabular}{|c|c|c|c|}
    \hline
    Voice&Acceleration&Light&Noise \\
    \hline
    180.9 KB & 91.7 KB & 4.4 KB & 5.4 KB  \\
    \hline
    \end{tabular}
    \label{File_size}
\end{table}

\section{Discussion}
\subsection{Smarter Triggering Algorithm}
The environment detection algorithms proposed in Section 4 intend to make the attack more stealthy. The proposed attack can be further improved to be more energy efficient. Once the activation voice is synthesised, the environment detection algorithm keeps collecting data from microphone, light sensor, and accelerometer. As we discussed in Section 5, though the power consumption is very low, it still has chance to be noticed by the user if it keeps running for a long time. We observed that in the situations that the environment detection module decides to launch the attack, the noise levels are higher than a certain threshold in most cases. As a result, the noise level can be a preliminary indicator for the detection algorithm. On standby mode, only microphone is activated to collect noise data, and once the noise level exceed a certain threshold, the other two sensors then start to collect data. We have tested that the power consumption can be further reduced from 0.8 mAh to 0.4 mAh, which makes it even harder to be noticed by the user.

\subsection{Essential Factors of the Attacking Success}

\textbf{Success in delivery.} The abuse of high risk permissions in Android apps making users insensitive to granting these permissions. For example, with the rise in popularity of location-based apps and Augmented Reality (AR) games, user are becoming less cautious when granting the permission to access the location and the camera, both of which are high risk permissions that may leak your private information. In our proof-of-concept attack, we disguise the spyware as a microphone controlled game to fool the user for granting the permission to access microphone.
\textbf{Success in avoid being detected by anti-malware products.} Currently, commercial anti-malware tools are based on known features of malware, such as signatures and sensitive operations. In our prototype spyware, there are no relevant signatures in existing signature library of anti-malware tools. All the sensitive operations in the attack, such sending emails and making phone calls, are executed through the Voice Assistant, which has privilege to access sensitive data, and not being monitored by the anti-malware products. 

\textbf{To avoid being noticed by users.} In the proposed attack, we developed a stealthy attacking module, which monitors the environment and looks for good time to launch the attack. It also adjusts the volume of the voice commands, to ensure that the voice commands can be captured and recognised by the smartphone, but cannot be heard by users.

\subsection{Defence Approaches}
\textbf{Identify the source of the voice commands.} In the proposed attack scenario, the voice commands are played via the built-in speaker on the smartphone. New techniques \cite{VoicePrint2017} are able to locate the source of the sound, which then can determine whether the sound comes from the built-in speaker. The voice assistant vendors can disable the proposed attack by setting the voice assistant as not to receive voice commands from the built-in speaker on its hosting smartphone. However, this defence approach requires additional devices to be worn, making it unrealistic for daily use.

\textbf{Continuous authentication for voice assistants.}
\cite{VAJudge2017} proposed a scheme that collects the body-surface vibrations of the user and matches it with the speech signal received by the voice assistant's microphone. The voice assistant executes only the commands that originate from the voice of the owner. While it may successfully defend the proposed attack, it also brings some inconvenience to the user, for example, the user cannot activate the voice assistant when they didn't hold the smartphone. Actually, users tend to interact with voice assistant when they are not able to touch the screen, such as when they are driving. 

\textbf{Distinguish human voice from machine-based voice}.  \cite{AttackDefence2017} explores the difference between a human voice and machine-based voice based on the magnetic field emitted from loudspeakers, which is able to detect machine-based voice impersonation attacks. However, false positive may be produced when there are other devices around, which generates magnetic signals. 

\subsection{Lessons from This Work}
In the newest Android OS version, there is a vulnerability of Google Assistant. Once the Google Assistant is activated, it can unlock the screen, change smartphone settings, and do a lot of operations that a human can do, such as sending SMS/emails and making phone calls. Due to the privilege it has to access system resources and private information, the Google Assistant can be a stepping stone for the attackers to hack into the smartphones. Though the users can turn off the “unlock with voice match” option, which requires the user to unlock the smartphone with PIN or fingerprints before the Google Assistant can be launched, it is against the "hands free" design target of the voice assistant. More secure mechanisms should be embedded to improve the security of the voice assistant, both from the research community and the OS vendor. 

\section{Related Work}
\subsection{Attack to Voice assistant}
To date, there are some methods that are designed to hack voice assistant. For example, Diao et al.\cite{YourVA2014} proposed an attacking method based on built-in speakers, which collected the surrounding light level as well as users' movement in early morning. They utilised the vulnerable inter-component communication `Intent (API: \textsc{\textit{ACTION\_VOICE\_SEARCH\_HANDS\_FREE}})' \cite{Intent} to launch attacks, in which a recorded voice command will be played to control the smart phones. Similar attacking methods are used in Efthimios and Constantinos's work \cite{Monkey2017}, where a linked device to transfer voice feedback. However, for the works \cite{YourVA2014,Monkey2017} the vulnerable API can only activate Google Assistant. It is not available for other voice assistants such as Samsung Bixby \cite{Bixby2018} and Cortana\cite{Cortana2014}.

%Efthimios and Constantinos\cite{Monkey2017} present a attacking method by using the linked device to transfer voice feedback. In this method, they also use intent to launch voice assistant, and fetch commands from Firebase, then use TTS to play the commands. This method has a dynamic attacking commands fetch but still missing the intelligent attack control. It has a narrow attack area when there are not linked device. This attack will be likely notice in the real world.

There are also some researchers studied making special audio to launch attacks. For example, Nicholas et al. \cite{HiddenVC2016} proposed a method to manipulate attacking audio. Their idea was to obfuscate the raw audio and make it sound like a noise. Technically, human beings might feel confused at the noise, but voice assistants embedded in smartphones could still precisely recognise the meaning of the manipulated audio. In another example, Zhang et al.\cite{DolphinAttack2017} presented a method to translate a TTS-based voice into an ultrasonic audio. They then directly played this inaudible audio to hack voice assistants. In fact, smartphones can receive a high-frequency sound waves, which cannot be heard by human beings (\textit{e.g.} audio frequency$\geq20$kHz). Therefore, attackers can use the ultrasonic audio commands to control the voice assistant. This attacking method needs an ultrasonic audio generator, which may not be practical in many real world scenarios. In particular, their method did not analyse the optimal attacking time. Therefore, the targeting user might be very vigilant when unsolicited window pop-up happened during the attacks. 

%it is similar to ultrasonic voice attack, even human can not notice the "noise", they can realise that the voice assistant has been launch. In additional, this method is lack of infection approach. Thus, this attack cannot launched in the real world.

To implement a more realistic and powerful voice attack, we build this stealthy spyware. The core differences are: 1) we use a `smart' algorithm to launch voice assistant in the daily use of mobile phone; the proposed attacking method is simple and can be applied to all types of voice assistants; 2) our attacking is more stealthy as the developed spyware will detect and determine when is the optimal attacking time.

\subsection{Context-Awareness.} 
It has become ubiquitous for various sensors in our life. People can utilise multi-sensors to predict human being's activities \cite{Sensor2009}. For example, there is a multi-sensor based human activity detection method for smart homes \cite{Sensor2009,Sensor2006,AR2009}. They used wearable sensors to collect data and adopted Coupled Hidden Markov Models (CHMMs) to recognise multi-user activities from sensor readings in a smart home environment. The main limitation of these systems was that they provided activity information only when the subject interacted with one of the tagged objects. Therefore, the only recognisable activities were those that involved these objects. Wiese et al.\cite{ContextA2013} presented an experiment to collect the sensor data to analyse where people keep their smartphones. They utilised different combinations of sensors embedded in most smartphones like accelerometer, proximity sensor, and light sensor to recognise the places of smartphones. According to their research, there were $85\%$ success rate to determine if a phone was in a bag, in a pocket, out, or in hand. Liu et al.\cite{SmartWatch2015} proposed the sensor-based method to recognise PINs when they were input by keyboard to smartwatch. They mainly used accelerometer to recognise the PINs and typed texts. They presented a set of new techniques to model user's hand movement and reduce the interference from noises. This attack could achieve high accuracy in keystroke inference.

\section{Conclusion}
In this paper, we proposed a stealthy attacking method targeting Voice Assistant on smartphones. With the proposed attacking method, the attacker can activate the Voice Assistant and apply further attacks, such as leaking private information, sending forged SMS/emails, and calling arbitrary numbers. An Intelligent Environment Detection module is designed to choose a optimal attacking time, thus can hide the attack from being noticed by users. Through our proof-of-concept attack targeting Google Assistant on Android platform, we demonstrated the feasibility of the attack in real-world scenarios.  This research may inspire the researchers and OS vendors rethink the security of Voice Assistant.
%end not revised 8
%\section*{Acknoledgement}

%The conference seeks submissions presenting novel research results in all aspects of computer and communications security and privacy, including both practical and theoretical contributions.

%\begin{acks}
% TODO: For the submission, don't include acknowledgments since they would most likely deanonymize you.
%\end{acks}
 % TODO: replace with your brilliant paper!

\bibliographystyle{ACM-Reference-Format}
\bibliography{ccs-sample}


\begin{thebibliography}{00}

%%% the Web address.

\ifx \showCODEN    \undefined \def \showCODEN     #1{\unskip}     \fi
\ifx \showDOI      \undefined \def \showDOI       #1{#1}\fi
\ifx \showISBNx    \undefined \def \showISBNx     #1{\unskip}     \fi
\ifx \showISBNxiii \undefined \def \showISBNxiii  #1{\unskip}     \fi
\ifx \showISSN     \undefined \def \showISSN      #1{\unskip}     \fi
\ifx \showLCCN     \undefined \def \showLCCN      #1{\unskip}     \fi
\ifx \shownote     \undefined \def \shownote      #1{#1}          \fi
\ifx \showarticletitle \undefined \def \showarticletitle #1{#1}   \fi
\ifx \showURL      \undefined \def \showURL       {\relax}        \fi
% invisible to TeX
\providecommand\bibfield[2]{#2}
\providecommand\bibinfo[2]{#2}
\providecommand\natexlab[1]{#1}
\providecommand\showeprint[2][]{arXiv:#2}

\bibitem[\protect\citeauthoryear{Alepis and Patsakis}{Alepis and
  Patsakis}{2017}]%
        {Monkey2017}
\bibfield{author}{\bibinfo{person}{Efthimios Alepis} {and}
  \bibinfo{person}{Constantinos Patsakis}.} \bibinfo{year}{2017}\natexlab{}.
\newblock \showarticletitle{Monkey Says, Monkey Does: Security and Privacy on
  Voice Assistants}.
\newblock \bibinfo{journal}{{\em IEEE Access\/}}  \bibinfo{volume}{5}
  (\bibinfo{year}{2017}), \bibinfo{pages}{17841--17851}.
\newblock


\bibitem[\protect\citeauthoryear{Android Developer}{Android Developer}{2018a}]%
        {AudioRecord}
Android Developer \bibinfo{year}{2018}\natexlab{a}.
\newblock \bibinfo{booktitle}{{\em AudioRecord}}.
\newblock Android Developer.
\newblock
\newblock
\shownote{\url{https://developer.android.com/reference/android/media/AudioRecord}.}


\bibitem[\protect\citeauthoryear{Android Developer}{Android Developer}{2018b}]%
        {Intent}
Android Developer \bibinfo{year}{2018}\natexlab{b}.
\newblock \bibinfo{booktitle}{{\em Intent}}.
\newblock Android Developer.
\newblock
\newblock
\shownote{\url{https://developer.android.com/reference/android/speech/RecognizerIntent.html}.}


\bibitem[\protect\citeauthoryear{Android Developer}{Android Developer}{2018c}]%
        {MediaRecorder}
Android Developer \bibinfo{year}{2018}\natexlab{c}.
\newblock \bibinfo{booktitle}{{\em MediaRecorder}}.
\newblock Android Developer.
\newblock
\newblock
\shownote{\url{https://developer.android.com/reference/android/media/MediaRecorder}.}


\bibitem[\protect\citeauthoryear{Android Developer}{Android Developer}{2018d}]%
        {SensorEventListener}
Android Developer \bibinfo{year}{2018}\natexlab{d}.
\newblock \bibinfo{booktitle}{{\em SensorEventListener}}.
\newblock Android Developer.
\newblock
\newblock
\shownote{\url{https://developer.android.com/reference/android/hardware/SensorEventListener}.}


\bibitem[\protect\citeauthoryear{Aron}{Aron}{2011}]%
        {Siri2011}
\bibfield{author}{\bibinfo{person}{Jacob Aron}.}
  \bibinfo{year}{2011}\natexlab{}.
\newblock \bibinfo{title}{How innovative is Apple's new voice assistant, Siri?}
\newblock   (\bibinfo{year}{2011}).
\newblock


\bibitem[\protect\citeauthoryear{Arp, Spreitzenbarth, H{\"u}bner, Gascon, and
  Rieck}{Arp et~al\mbox{.}}{2014}]%
        {Drebin2014}
\bibfield{author}{\bibinfo{person}{Daniel Arp}, \bibinfo{person}{Michael
  Spreitzenbarth}, \bibinfo{person}{Malte H{\"u}bner}, \bibinfo{person}{Hugo
  Gascon}, {and} \bibinfo{person}{Konrad Rieck}.}
  \bibinfo{year}{2014}\natexlab{}.
\newblock \showarticletitle{Drebin: Efficient and Explainable Detection of
  Android Malware in Your Pocket}.
\newblock


\bibitem[\protect\citeauthoryear{Canbek and Mutlu}{Canbek and Mutlu}{2016}]%
        {PersonalAssistant2016}
\bibfield{author}{\bibinfo{person}{Nil~Goksel Canbek} {and}
  \bibinfo{person}{Mehmet~Emin Mutlu}.} \bibinfo{year}{2016}\natexlab{}.
\newblock \showarticletitle{On the track of Artificial Intelligence: Learning
  with intelligent personal assistants}.
\newblock \bibinfo{journal}{{\em Journal of Human Sciences\/}}
  \bibinfo{volume}{13}, \bibinfo{number}{1} (\bibinfo{year}{2016}),
  \bibinfo{pages}{592--601}.
\newblock


\bibitem[\protect\citeauthoryear{Carlini, Mishra, Vaidya, Zhang, Sherr,
  Shields, Wagner, and Zhou}{Carlini et~al\mbox{.}}{2016}]%
        {HiddenVC2016}
\bibfield{author}{\bibinfo{person}{Nicholas Carlini}, \bibinfo{person}{Pratyush
  Mishra}, \bibinfo{person}{Tavish Vaidya}, \bibinfo{person}{Yuankai Zhang},
  \bibinfo{person}{Micah Sherr}, \bibinfo{person}{Clay Shields},
  \bibinfo{person}{David~A. Wagner}, {and} \bibinfo{person}{Wenchao Zhou}.}
  \bibinfo{year}{2016}\natexlab{}.
\newblock \showarticletitle{Hidden Voice Commands}. In \bibinfo{booktitle}{{\em
  USENIX Security Symposium}}.
\newblock


\bibitem[\protect\citeauthoryear{Chen, Ren, Piao, Wang, Wang, Weng, Su, and
  Mohaisen}{Chen et~al\mbox{.}}{2017}]%
        {AttackDefence2017}
\bibfield{author}{\bibinfo{person}{Si Chen}, \bibinfo{person}{Kui Ren},
  \bibinfo{person}{Sixu Piao}, \bibinfo{person}{Cong Wang},
  \bibinfo{person}{Qian Wang}, \bibinfo{person}{Jian Weng}, \bibinfo{person}{Lu
  Su}, {and} \bibinfo{person}{Aziz Mohaisen}.} \bibinfo{year}{2017}\natexlab{}.
\newblock \showarticletitle{You Can Hear But You Cannot Steal: Defending
  Against Voice Impersonation Attacks on Smartphones}.
\newblock \bibinfo{journal}{{\em 2017 IEEE 37th International Conference on
  Distributed Computing Systems (ICDCS)\/}} (\bibinfo{year}{2017}),
  \bibinfo{pages}{183--195}.
\newblock


\bibitem[\protect\citeauthoryear{De~Silva}{De~Silva}{2009}]%
        {Sensor2009}
\bibfield{author}{\bibinfo{person}{Liyanage~C. De~Silva}.}
  \bibinfo{year}{2009}\natexlab{}.
\newblock \showarticletitle{Multi-sensor Based Human Activity Detection for
  Smart Homes}. In \bibinfo{booktitle}{{\em Proceedings of the 3rd
  International Universal Communication Symposium}} {\em (\bibinfo{series}{IUCS
  '09})}. \bibinfo{pages}{223--229}.
\newblock
\showISBNx{978-1-60558-641-0}


\bibitem[\protect\citeauthoryear{Diao, Liu, Zhou, and Zhang}{Diao
  et~al\mbox{.}}{2014}]%
        {YourVA2014}
\bibfield{author}{\bibinfo{person}{Wenrui Diao}, \bibinfo{person}{Xiangyu Liu},
  \bibinfo{person}{Zhe Zhou}, {and} \bibinfo{person}{Kehuan Zhang}.}
  \bibinfo{year}{2014}\natexlab{}.
\newblock \showarticletitle{Your Voice Assistant is Mine: How to Abuse Speakers
  to Steal Information and Control Your Phone}. In \bibinfo{booktitle}{{\em
  Proceedings of the 4th ACM Workshop on Security and Privacy in Smartphones
  \&\#38; Mobile Devices}} {\em (\bibinfo{series}{SPSM '14})}.
  \bibinfo{pages}{63--74}.
\newblock
\showISBNx{978-1-4503-3155-5}


\bibitem[\protect\citeauthoryear{Dietterich and Horvitz}{Dietterich and
  Horvitz}{2015}]%
        {AI2015}
\bibfield{author}{\bibinfo{person}{Thomas~G Dietterich} {and}
  \bibinfo{person}{Eric~J Horvitz}.} \bibinfo{year}{2015}\natexlab{}.
\newblock \showarticletitle{Rise of concerns about AI: reflections and
  directions}.
\newblock \bibinfo{journal}{{\it Commun. ACM}} \bibinfo{volume}{58},
  \bibinfo{number}{10} (\bibinfo{year}{2015}), \bibinfo{pages}{38--40}.
\newblock


\bibitem[\protect\citeauthoryear{Feng, Fawaz, and Shin}{Feng
  et~al\mbox{.}}{2017}]%
        {VAJudge2017}
\bibfield{author}{\bibinfo{person}{Huan Feng}, \bibinfo{person}{Kassem Fawaz},
  {and} \bibinfo{person}{Kang~G. Shin}.} \bibinfo{year}{2017}\natexlab{}.
\newblock \showarticletitle{Continuous Authentication for Voice Assistants}. In
  \bibinfo{booktitle}{{\em Proceedings of the 23rd Annual International
  Conference on Mobile Computing and Networking}} {\em
  (\bibinfo{series}{MobiCom '17})}. \bibinfo{pages}{343--355}.
\newblock
\showISBNx{978-1-4503-4916-1}


\bibitem[\protect\citeauthoryear{Googl}{Googl}{2018}]%
        {VAFunctions}
Googl \bibinfo{year}{2018}\natexlab{}.
\newblock \bibinfo{booktitle}{{\em What can your Google Assistant do}}.
\newblock Googl.
\newblock
\newblock
\shownote{\url{https://assistant.google.com/explore?hl=en-AU}.}


\bibitem[\protect\citeauthoryear{Hawking, Russell, Tegmark, and
  Wilczek}{Hawking et~al\mbox{.}}{2014}]%
        {Cortana2014}
\bibfield{author}{\bibinfo{person}{Stephen Hawking}, \bibinfo{person}{Stuart
  Russell}, \bibinfo{person}{Max Tegmark}, {and} \bibinfo{person}{Frank
  Wilczek}.} \bibinfo{year}{2014}\natexlab{}.
\newblock \showarticletitle{Stephen Hawking:$\backslash$'Transcendence looks at
  the implications of artificial intelligence-but are we taking AI seriously
  enough?$\backslash$'}.
\newblock \bibinfo{journal}{{\em The Independent\/}} \bibinfo{volume}{2014},
  \bibinfo{number}{05-01} (\bibinfo{year}{2014}), \bibinfo{pages}{9313474}.
\newblock


\bibitem[\protect\citeauthoryear{Jiang, Hassan~Awadallah, Jones, Ozertem,
  Zitouni, Gurunath~Kulkarni, and Khan}{Jiang et~al\mbox{.}}{2015}]%
        {VAEvaluation2015}
\bibfield{author}{\bibinfo{person}{Jiepu Jiang}, \bibinfo{person}{Ahmed
  Hassan~Awadallah}, \bibinfo{person}{Rosie Jones}, \bibinfo{person}{Umut
  Ozertem}, \bibinfo{person}{Imed Zitouni}, \bibinfo{person}{Ranjitha
  Gurunath~Kulkarni}, {and} \bibinfo{person}{Omar~Zia Khan}.}
  \bibinfo{year}{2015}\natexlab{}.
\newblock \showarticletitle{Automatic Online Evaluation of Intelligent
  Assistants}. In \bibinfo{booktitle}{{\em Proceedings of the 24th
  International Conference on World Wide Web}} {\em (\bibinfo{series}{WWW
  '15})}. \bibinfo{pages}{506--516}.
\newblock
\showISBNx{978-1-4503-3469-3}


\bibitem[\protect\citeauthoryear{Kasmi and Esteves}{Kasmi and Esteves}{2015}]%
        {IEMITF2015}
\bibfield{author}{\bibinfo{person}{Chaouki Kasmi} {and}
  \bibinfo{person}{Jose~Lopes Esteves}.} \bibinfo{year}{2015}\natexlab{}.
\newblock \showarticletitle{IEMI Threats for Information Security: Remote
  Command Injection on Modern Smartphones}.
\newblock \bibinfo{journal}{{\em IEEE Transactions on Electromagnetic
  Compatibility\/}}  \bibinfo{volume}{57} (\bibinfo{year}{2015}),
  \bibinfo{pages}{1752--1755}.
\newblock


\bibitem[\protect\citeauthoryear{Kiseleva, Williams, Jiang, Hassan~Awadallah,
  Crook, Zitouni, and Anastasakos}{Kiseleva et~al\mbox{.}}{2016}]%
        {VAUseage2016}
\bibfield{author}{\bibinfo{person}{Julia Kiseleva}, \bibinfo{person}{Kyle
  Williams}, \bibinfo{person}{Jiepu Jiang}, \bibinfo{person}{Ahmed
  Hassan~Awadallah}, \bibinfo{person}{Aidan~C. Crook}, \bibinfo{person}{Imed
  Zitouni}, {and} \bibinfo{person}{Tasos Anastasakos}.}
  \bibinfo{year}{2016}\natexlab{}.
\newblock \showarticletitle{Understanding User Satisfaction with Intelligent
  Assistants}. In \bibinfo{booktitle}{{\em Proceedings of the 2016 ACM on
  Conference on Human Information Interaction and Retrieval}} {\em
  (\bibinfo{series}{CHIIR '16})}. \bibinfo{pages}{121--130}.
\newblock
\showISBNx{978-1-4503-3751-9}


\bibitem[\protect\citeauthoryear{Knote, Janson, Eigenbrod, and
  S{\"o}llner}{Knote et~al\mbox{.}}{2018}]%
        {Bixby2018}
\bibfield{author}{\bibinfo{person}{Robin Knote}, \bibinfo{person}{Andreas
  Janson}, \bibinfo{person}{Laura Eigenbrod}, {and} \bibinfo{person}{Matthias
  S{\"o}llner}.} \bibinfo{year}{2018}\natexlab{}.
\newblock \showarticletitle{The What and How of Smart Personal Assistants:
  Principles and Application Domains for IS Research}.
\newblock  (\bibinfo{year}{2018}).
\newblock


\bibitem[\protect\citeauthoryear{Liu, Cornelius, Rawassizadeh, Peterson, and
  Kotz}{Liu et~al\mbox{.}}{2017}]%
        {VoicePrint2017}
\bibfield{author}{\bibinfo{person}{Rui Liu}, \bibinfo{person}{Cory Cornelius},
  \bibinfo{person}{Reza Rawassizadeh}, \bibinfo{person}{Ron Peterson}, {and}
  \bibinfo{person}{David Kotz}.} \bibinfo{year}{2017}\natexlab{}.
\newblock \showarticletitle{Poster: Vocal Resonance As a Passive Biometric}. In
  \bibinfo{booktitle}{{\em Proceedings of the 15th Annual International
  Conference on Mobile Systems, Applications, and Services}} {\em
  (\bibinfo{series}{MobiSys '17})}. \bibinfo{pages}{160--160}.
\newblock
\showISBNx{978-1-4503-4928-4}


\bibitem[\protect\citeauthoryear{Liu, Zhou, Diao, Li, and Zhang}{Liu
  et~al\mbox{.}}{2015}]%
        {SmartWatch2015}
\bibfield{author}{\bibinfo{person}{Xiangyu Liu}, \bibinfo{person}{Zhe Zhou},
  \bibinfo{person}{Wenrui Diao}, \bibinfo{person}{Zhou Li}, {and}
  \bibinfo{person}{Kehuan Zhang}.} \bibinfo{year}{2015}\natexlab{}.
\newblock \showarticletitle{When Good Becomes Evil: Keystroke Inference with
  Smartwatch}. In \bibinfo{booktitle}{{\em Proceedings of the 22Nd ACM SIGSAC
  Conference on Computer and Communications Security}} {\em
  (\bibinfo{series}{CCS '15})}. \bibinfo{pages}{1273--1285}.
\newblock
\showISBNx{978-1-4503-3832-5}


\bibitem[\protect\citeauthoryear{Pedregosa, Varoquaux, Gramfort, Michel,
  Thirion, Grisel, Blondel, Prettenhofer, Weiss, Dubourg,
  et~al\mbox{.}}{Pedregosa et~al\mbox{.}}{2011}]%
        {Scikit2011}
\bibfield{author}{\bibinfo{person}{Fabian Pedregosa}, \bibinfo{person}{Ga{\"e}l
  Varoquaux}, \bibinfo{person}{Alexandre Gramfort}, \bibinfo{person}{Vincent
  Michel}, \bibinfo{person}{Bertrand Thirion}, \bibinfo{person}{Olivier
  Grisel}, \bibinfo{person}{Mathieu Blondel}, \bibinfo{person}{Peter
  Prettenhofer}, \bibinfo{person}{Ron Weiss}, \bibinfo{person}{Vincent
  Dubourg}, {et~al\mbox{.}}} \bibinfo{year}{2011}\natexlab{}.
\newblock \showarticletitle{Scikit-learn: Machine learning in Python}.
\newblock \bibinfo{journal}{{\em Journal of machine learning research\/}}
  \bibinfo{volume}{12}, \bibinfo{number}{Oct} (\bibinfo{year}{2011}),
  \bibinfo{pages}{2825--2830}.
\newblock


\bibitem[\protect\citeauthoryear{Pirttikangas, Fujinami, and
  Nakajima}{Pirttikangas et~al\mbox{.}}{2006}]%
        {Sensor2006}
\bibfield{author}{\bibinfo{person}{Susanna Pirttikangas},
  \bibinfo{person}{Kaori Fujinami}, {and} \bibinfo{person}{Tatsuo Nakajima}.}
  \bibinfo{year}{2006}\natexlab{}.
\newblock \showarticletitle{Feature Selection and Activity Recognition from
  Wearable Sensors}. In \bibinfo{booktitle}{{\em Proceedings of the Third
  International Conference on Ubiquitous Computing Systems}} {\em
  (\bibinfo{series}{UCS'06})}. \bibinfo{pages}{516--527}.
\newblock
\showISBNx{3-540-46287-2, 978-3-540-46287-3}


\bibitem[\protect\citeauthoryear{Royal, Halpin, Dagon, Edmonds, and Lee}{Royal
  et~al\mbox{.}}{2006}]%
        {HidenCode2006}
\bibfield{author}{\bibinfo{person}{Paul Royal}, \bibinfo{person}{Mitch Halpin},
  \bibinfo{person}{David Dagon}, \bibinfo{person}{Robert Edmonds}, {and}
  \bibinfo{person}{Wenke Lee}.} \bibinfo{year}{2006}\natexlab{}.
\newblock \showarticletitle{PolyUnpack: Automating the Hidden-Code Extraction
  of Unpack-Executing Malware.}. In \bibinfo{booktitle}{{\em ACSAC}}.
  \bibinfo{publisher}{IEEE Computer Society}, \bibinfo{pages}{289--300}.
\newblock
\showISBNx{0-7695-2716-7}


\bibitem[\protect\citeauthoryear{Wang, Gu, Tao, and Lu}{Wang
  et~al\mbox{.}}{2009}]%
        {AR2009}
\bibfield{author}{\bibinfo{person}{Liang Wang}, \bibinfo{person}{Tao Gu},
  \bibinfo{person}{Xianping Tao}, {and} \bibinfo{person}{Jian Lu}.}
  \bibinfo{year}{2009}\natexlab{}.
\newblock \showarticletitle{Sensor-Based Human Activity Recognition in a
  Multi-user Scenario}. In \bibinfo{booktitle}{{\em Proceedings of the European
  Conference on Ambient Intelligence}} {\em (\bibinfo{series}{AmI '09})}.
  \bibinfo{pages}{78--87}.
\newblock
\showISBNx{978-3-642-05407-5}


\bibitem[\protect\citeauthoryear{Wiese, Saponas, and Brush}{Wiese
  et~al\mbox{.}}{2013}]%
        {ContextA2013}
\bibfield{author}{\bibinfo{person}{Jason Wiese}, \bibinfo{person}{T.~Scott
  Saponas}, {and} \bibinfo{person}{A.J.~Bernheim Brush}.}
  \bibinfo{year}{2013}\natexlab{}.
\newblock \showarticletitle{Phoneprioception: Enabling Mobile Phones to Infer
  Where They Are Kept}. In \bibinfo{booktitle}{{\em Proceedings of the SIGCHI
  Conference on Human Factors in Computing Systems}} {\em (\bibinfo{series}{CHI
  '13})}. \bibinfo{pages}{2157--2166}.
\newblock
\showISBNx{978-1-4503-1899-0}


\bibitem[\protect\citeauthoryear{Xu, Zhang, Luo, Jia, Xuan, and Teng}{Xu
  et~al\mbox{.}}{2009}]%
        {XN2009}
\bibfield{author}{\bibinfo{person}{Nan Xu}, \bibinfo{person}{Fan Zhang},
  \bibinfo{person}{Yisha Luo}, \bibinfo{person}{Weijia Jia},
  \bibinfo{person}{Dong Xuan}, {and} \bibinfo{person}{Jin Teng}.}
  \bibinfo{year}{2009}\natexlab{}.
\newblock \showarticletitle{Stealthy Video Capturer: A New Video-based Spyware
  in 3G Smartphones}. In \bibinfo{booktitle}{{\em Proceedings of the Second ACM
  Conference on Wireless Network Security}} {\em (\bibinfo{series}{WiSec
  '09})}. \bibinfo{pages}{69--78}.
\newblock


\bibitem[\protect\citeauthoryear{Zhang, Yan, Ji, Zhang, Zhang, and Xu}{Zhang
  et~al\mbox{.}}{2017}]%
        {DolphinAttack2017}
\bibfield{author}{\bibinfo{person}{Guoming Zhang}, \bibinfo{person}{Chen Yan},
  \bibinfo{person}{Xiaoyu Ji}, \bibinfo{person}{Tianchen Zhang},
  \bibinfo{person}{Taimin Zhang}, {and} \bibinfo{person}{Wenyuan Xu}.}
  \bibinfo{year}{2017}\natexlab{}.
\newblock \showarticletitle{DolphinAttack: Inaudible Voice Commands}. In
  \bibinfo{booktitle}{{\em Proceedings of the 2017 ACM SIGSAC Conference on
  Computer and Communications Security}} {\em (\bibinfo{series}{CCS '17})}. 15.
\newblock


\end{thebibliography}

\end{document}